\documentclass[%
 reprint,
nofootinbib,
 amsmath,amssymb,
 aps,
]{revtex4-1}

\usepackage{graphicx}
\usepackage{dcolumn}
\usepackage{bm}
\usepackage{longtable}%

\begin{document}

\preprint{APS/123-QED}

\title{Relativistic effects due to gravimagnetic moment of a rotating body}

\author{Walberto Guzm\'an Ram\'irez$^1$}

\email{wguzman@cbpf.br}

\author{Alexei A. Deriglazov$^{1,2}$}%

\email{alexei.deriglazov@ufjf.edu.br}
 

\affiliation{$^1$Depto. de Matem\'atica, ICE, Universidade Federal de Juiz de Fora, MG, Brazil \\}

\affiliation{$^2$Laboratory of Mathematical Physics, Tomsk Polytechnic University, \\ 634050 Tomsk, Lenin Ave. 30,
Russian Federation
}%




\date{\today}

\begin{abstract}
We compute exact Hamiltonian (and corresponding Dirac brackets) for spinning particle with gravimagnetic moment
$\kappa$ in an arbitrary gravitational background. $\kappa=0$ corresponds to the Mathisson-Papapetrou-Tulczyjew-Dixon
(MPTD) equations. $\kappa=1$ leads to modified MPTD equations with reasonable behavior in the ultrarelativistic limit. So
we study the modified equations in the leading post-Newtonian approximation. Rotating body with unit gravimagnetic
moment has qualitatively different behavior as compared with MPTD body: A) If a number of gyroscopes with various
rotation axes are freely traveling together, the angles between the axes change with time. B) For specific binary
systems, gravimagnetic moment gives a contribution to frame-dragging effect with the magnitude, that turns out to be
comparable with that of Schiff frame dragging. 
\begin{description}
\item[PACS numbers] 04.20.-q, 03.65.Sq, 04.20.Fy, 04.20.Cv, 04.80.Cc.
\end{description}
\end{abstract}

\pacs{Valid PACS appear here}
\maketitle


\section{Introduction}

Rotating body in general relativity is usually described on the base of manifestly generally covariant
Mathisson-Papapetrou-Tulczyjew-Dixon (MPTD) equations, that prescribe the dynamics of both trajectory and spin of the
body in an external gravitational field \cite{Mathisson:1937zz, Fock1939, Papapetrou:1951pa, Tulc, Dixon1964,
pirani:1956}. Starting from the pioneer works, these equations were considered as a Hamiltonian-type system. In the
recent work \cite{DW2015.1}, we realized this idea by constructing the minimal interaction with gravity in the vector
model of spinning particle, and showed that this indeed leads to MPTD equations in the Hamiltonian formalism (see also
below). This allowed us to study ultra relativistic limit in exact equations for trajectory of MPTD particle in the
laboratory time.  Using the Landau-Lifshitz $(1+3)$\,-decomposition \cite{bib16} we  observed that, unlike a geodesic
equation, the MPTD equations lead to the expression for three-acceleration which contains divergent terms as
$v\rightarrow c$ \cite{DWGR2017}.  Fast test particles are now under intensive investigation \cite{Amorim:2010qj,
Abreu:2010mt, Bruno15, Kai1703, Kai1705}, and represent an important  tool in the study, for example,  of near horizon
geometry of black holes \cite{Khriplovich1989, Pomeransky1998, khriplovich:1996, Will2014, Gal2013,
 Dolan:2013roa, Akcay:2016dku, Bini:2017slb, Plyatsko:2016bee}. So, it would be
interesting to find a generalization of MPTD equations with  improved behavior in ultra relativistic regime. This can
be achieved, if we add a non-minimal spin-gravity interaction through gravimagnetic moment \cite{DWGR2016}. In the
theory with unit gravimagnetic moment, both acceleration and spin torque have reasonable behavior in ultra relativistic
limit. In the present work we study the modified equations in the regime of small velocities in the leading
post-Newtonian approximation. In Schwarzschild and Kerr space-times, the modified equations imply a number of
qualitatively new effects, that could be used to test experimentally, whether a rotating body in general relativity has
null or unit gravimagnetic moment.

The work is organized as follows. In Sect. \ref{sect.2} we shortly describe Lagrangian and Hamiltonian formulations of
vector model of spinning particle and compute Dirac brackets of the theory in an arbitrary gravitational background. In
the formulation with use of Dirac brackets, the complete Hamiltonian acquires a simple and expected form, while an
approximate $\frac{1}{c^2}$ Hamiltonian, further obtained in Sect. \ref{sect.4}, strongly resembles that of spinning
particle in electromagnetic background. This is in correspondence with the known analogy between gravity and
electromagnetism \cite{Thirring1918.1, Wald1972, Costa:2012cy, Natario:2007pu}. In Sect. \ref{sect.3} we introduce non-minimal spin-gravity
interaction through the gravimagnetic moment and obtain the corresponding equations of motion.  We show that constants
of motion due to isometries of space-time for the MPTD and the modified equations are the same. In section \ref{sect.4} we
compute the leading post-Newtonian corrections to the trajectory and spin of our particle with unit gravimagnetic
moment, and present the corresponding effective Hamiltonian in $\frac{1}{c^2}$\,-approximation. The non-minimal
interaction implies  extra contributions into both trajectory and spin, as compared with MPTD equations in the same
approximation. A number of effects due to non-minimal interaction are discussed in Sect. \ref{sect.5}.

{\bf Notation.} Our variables are taken in arbitrary parametrization $\tau$, then $\dot x^\mu=\frac{dx^\mu}{d\tau}$.
The square brackets mean antisymmetrization, $\omega^{[\mu}\pi^{\nu]}=\omega^\mu\pi^\nu-\omega^\nu\pi^\mu$. For the
four-dimensional quantities we suppress the contracted indexes and use the notation  $\dot x^\mu G_{\mu\nu}\dot
x^\nu=\dot xG\dot x$,  $N^\mu{}_\nu\dot x^\nu=(N\dot x)^\mu$, $\omega^2=g_{\mu\nu}\omega^\mu\omega^\nu$, $\mu, \nu=0,
1, 2, 3$.  Notation for the scalar functions constructed from second-rank tensors are $\theta S=
\theta^{\mu\nu}S_{\mu\nu}$, $S^2=S^{\mu\nu}S_{\mu\nu}$. When we work in four-dimensional Minkowski space with
coordinates $x^\mu=(x^0=ct,~  x^i)$, we use the metric $\eta_{\mu\nu}=(-, +, +, +)$, then $\dot x\omega=\dot
x^\mu\omega_\mu=-\dot x^0\omega^0+\dot x^i\omega^i$ and so on. Suppressing the indexes of three-dimensional quantities,
we use bold letters, $v^i\gamma_{ij}a^j={\bf v}\gamma{\bf a}$, $v^iG_{i\mu}v^\mu={\bf v}Gv$, $i, j=1, 2, 3$, and so on.

The covariant derivative is $\nabla
\omega^\mu=\frac{d\omega^\mu}{d\tau}+\Gamma^\mu_{\alpha\beta}\dot x^\alpha \omega^\beta$. The tensor of Riemann
curvature is $R^\sigma{}_{\lambda\mu\nu}=\partial_\mu\Gamma^\sigma{}_{\lambda\nu}
-\partial_\nu \Gamma^\sigma{}_{\lambda\mu}+\Gamma^\sigma{}_{\beta\mu}\Gamma^{\beta}{}_{\lambda\nu}-
\Gamma^\sigma{}_{\beta\nu}\Gamma^{\beta}{}_{\lambda\mu}$.


\section{Vector model of spin and Mathisson-Papapetrou-Tulczyjew-Dixon equations}\label{sect.2}

In the vector model of spin presented in \cite{deriglazov2014Monster}, the configuration space consist of the position of the particle
$x^\mu(\tau)$, and the vector $\omega^\mu(\tau)$ attached to the point $x^\mu(\tau)$. Minimal interaction with
gravity is achieved by direct covariantization of the free action, initially formulated in Minkowski space.
That is we replace $\eta_{\mu\nu}\rightarrow g_{\mu\nu}$, and usual derivative of the vector $\omega^\mu$ by the
covariant derivative: $\dot\omega^\mu\rightarrow\nabla\omega^\mu$. The resulting Lagrangian action reads \cite{DW2015.1}
\begin{eqnarray}\label{L-curved}
S =&& -\frac{1}{\sqrt{2}} \int d\tau \left[ m^2c^2 -\frac{\alpha}{\omega^2} \right]^{\frac 12}  \nonumber \\
&&\times \sqrt{-\dot x N \dot x -
\nabla\omega N \nabla\omega + T^{1/2}} \,.
\end{eqnarray}
We have denoted $T\equiv [\dot x N\dot x + \nabla\omega N \nabla\omega]^2- 4 (\dot x N \nabla\omega )^2$, and
$N_{\mu\nu}\equiv  g_{\mu\nu}-\frac{\omega_\mu \omega_\nu}{\omega^2}$. The matrix $N$ is a projector on the plane
orthogonal to $\omega$: $N_{\mu\nu}\omega^\nu=0$. The parameter $\alpha$ determines the value of spin, in particular,
$\alpha=\frac{3\hbar^2}{4}$ corresponds to the spin one-half particle. In the spinless limit, $\omega^\mu=0$ and
$\alpha=0$, Eq. (\ref{L-curved}) reduces to the standard Lagrangian of a point particle, $-mc\sqrt{-\dot x^2}$.

The action is manifestly invariant under general-coordinate transformations as well as under reparametrizations of the
evolution parameter $\tau$. Besides, there is one more local symmetry, which acts in spin-sector and called the
spin-plane symmetry: the action remains invariant under rotations of the vectors $\omega^\mu$ and
$\pi_\mu=\frac{\partial L}{\partial\dot\omega^\mu}$ in their own plane \cite{deriglazov2010classical}. Being affected
by the local transformation, these vectors do not represent observable quantities. But their combination,
$S^{\mu\nu}=2(\omega^\mu\pi^\nu-\omega^\nu\pi^\mu)$, is an invariant quantity, which represents the spin-tensor of the
particle. We decompose the spin-tensor as follows:
\begin{eqnarray}\label{FF.2}
S^{\mu\nu}=2(\omega^\mu\pi^\nu-\omega^\nu\pi^\mu)=(S^{i0}=D^i, S_{ij}=2\epsilon_{ijk}S_k),
\end{eqnarray}
where $S_i$ is three-dimensional spin-vector, and $D_i$ is dipole electric moment \cite{ba1}.

Since we deal with a local-invariant theory and, furthermore, one of the basic observables is constructed from the
phase-space variables, the Hamiltonian formalism is the most convenient for analyzing the dynamics of the theory. So,
we first obtain the Hamiltonian equations of motion, and next, excluding  momenta,  we arrive at the Lagrangian
equations for the physical-sector variables $x$ and $S$.

Conjugate momenta for $x^\mu$ and $\omega^\mu$ are $p_\mu=\frac{\partial L}{\partial\dot x^\mu}$ and
$\pi_\mu=\frac{\partial L}{\partial\dot\omega^\mu}$ respectively. Due to the presence of $\dot x^\mu$ in
$\nabla\omega^\mu$, the conjugated momentum $p_\mu$ does not transform as a vector, so it is convenient to define the
canonical momentum
\begin{equation}\label{canonical-P}
P_\mu\equiv p_\mu-\Gamma^\beta_{\alpha\mu}\omega^\alpha\pi_\beta \, ,
\end{equation}
which   transforms as a vector under general-coordinate transformations.
The full set of phase-space coordinates consists of the pairs $x^\mu, p_\mu$ and  $\omega^\mu, \pi_\mu$.  They fulfill
the fundamental Poisson  brackets
$\{x^\mu ,  p_\nu \}=\delta^\mu_{\nu}$,
$\{\omega^\mu , \pi_\nu\}=\delta^\mu_\nu$,
then
\begin{eqnarray}\label{PB-1.0}
\{P_\mu , \omega^\nu \}=\Gamma^\nu_{\mu\alpha}\omega^\alpha, \quad \{ P_\mu , \pi_\nu \}=- \Gamma^\alpha_{\mu\nu}
\pi_\alpha, \cr
 \{P_\mu , \omega^2\}=\{P_\mu , \pi^2\}=\{P_\mu , \omega\pi\}=0.
\end{eqnarray}
For the quantities $x^\mu$, $P^\mu$ and $S^{\mu\nu}$, the basic Poisson brackets imply the typical relations used by
people for spinning particles in Hamiltonian formalism
\begin{eqnarray}\label{br}
&&\{ x^\mu , P_\nu \} =\delta^\mu_\nu, \quad \{ P_\mu , P_\nu\}=-\frac14 R_{\mu\nu\alpha\beta}S^{\alpha\beta}, \nonumber \\
 &&\{P_\mu, S^{\alpha\beta} \}=\Gamma^{\alpha}_{\mu\sigma}S^{\sigma\beta}-\Gamma^\beta_{\mu\sigma}S^{\sigma\alpha} \, , \nonumber \\
&&\{S^{\mu\nu},S^{\alpha\beta}\}= 2(g^{\mu\alpha} S^{\nu\beta}-g^{\mu\beta} S^{\nu\alpha}-g^{\nu\alpha} S^{\mu\beta}
+g^{\nu\beta} S^{\mu\alpha})\,. \qquad \qquad \quad
\end{eqnarray}
Applying  the Dirac-Bergman procedure for a singular system to the theory (\ref{L-curved}), we arrive at the
Hamiltonian \cite{DWGR2017}
\begin{eqnarray}\label{Hamiltonian}
H= \frac{\lambda_1}{2} [T_1 + 4a (\pi \theta P) T_3 - 4a(\omega \theta P) T_4+T_5 ]
+ \lambda_2 T_2 \, ,
\end{eqnarray}
composed of the constraints
\begin{eqnarray}
&T_1\equiv P^2 + m^2 c^2  =0 ,   \label{primary}\\
&T_2\equiv\omega\pi=0, \quad  T_3 \equiv P\omega =0,  \quad T_4 \equiv P\pi =0,  \nonumber\\
&T_5\equiv \pi^2 - \frac{\alpha}{\omega^2}= 0 \,. \label{primaryb}
\end{eqnarray}
In the expression for $H$ we have denoted
\begin{equation}\label{theta}
\theta_{\mu\nu} \equiv R_{\alpha\beta\mu\nu}S^{\alpha\beta} \,, \qquad
a \equiv \frac{2}{16m^2c^2+(\theta S)} \,.
\end{equation}
The antisymmetric tensor $\theta_{\mu\nu}$ turns out to be gravitational analogy of the electromagnetic field strength
$F_{\mu\nu}$, see below. $T_1, \ldots, T_4$ appear as the primary constraints in the course of Dirac-Bergmann
procedure,  $T_5$ is the only secondary constraint of the theory, and $\lambda_1, \lambda_2$ are the Lagrangian
multipliers associated to $T_1$ and $T_2$.  Poisson brackets of the constraints are summarized in Table
\ref{algebra-constraints}. The Table implies that $T_3$ and $T_4$ represent a pair of second-class constraints, while
$T_2$, $T_5$ and the  combination $T_1 + 4a (\pi \theta P) T_3 - 4a(\omega \theta P) T_4$ are  the first-class
constraints.  So the Hamiltonian (\ref{Hamiltonian}) consist of the first-class constraints.


\begin{table*}
\caption{Poisson brackets of constraints}\label{algebra-constraints}
\begin{tabular}{|c|c|c|c|c|c|}
\hline
                              & $\qquad T_1 \qquad$  & $T_5$         & $T_2$          & $T_3$                 & $T_4$       \\  \hline \hline
$T_1=P^2+m^2 c^2$         & 0             & 0             & 0              &$ \frac 12 (\omega\theta P)$                    &$ \frac 12 (\pi\theta P)$     \\

                              &               &               &           &            &              \\ \hline
$ T_5=\pi^2 -\frac{\alpha}{\omega^2}    $               & 0             &   0   &  $-2T_5$             & $-2T_4$   & $-2\alpha T_3/(\omega^2)^2$   \\
& ${}$ &      &           &                &      \\
\hline
$T_2=\omega\pi$            & $0$             & $2T_5$  & $0$            & $-T_3$          & $T_4$     \\
& ${}$ &      &           &                &        \\   \hline
$T_3=P\omega$               & $-\frac 12 (\omega\theta P)$    & $2T_4$      & $T_3$   &     0                 & $P^2-\frac{(\theta S)}{16}\approx-\frac{1}{8a}$  \\
& ${}$ &      &           &                &        \\  \hline
$T_4=P \pi$       & $-\frac 12 (\pi\theta P)$  &$2\alpha T_3/(\omega^2)^2$  & $-T_4$&   $-P^2+\frac{(\theta S)}{16}\approx\frac{1}{8a}$               & 0        \\
                              & ${}$ &      &           &                &     \\   \hline
\end{tabular}
\end{table*}
%


Taking into account that each second-class constraint rules out one phase-space variable, whereas each first-class
constraint rules out two variables, we have the right number of spin degrees of freedom, $8-(2+4)=2$. The meaning of
the constraints becomes clear if we consider their effect over the spin tensor. The second-class constraints $T_3=0$
and $T_4=0$ imply the spin supplementary condition
\begin{equation}
S^{\mu\nu}P_\nu = 0 \, ,\label{condition1}
\end{equation}
while the first-class constraints $T_2$ and $T_5$ fix the value of square of the spin tensor
\begin{equation}
S^{\mu\nu} S_{\mu\nu} = 8\alpha. \label{condition2}
\end{equation}
The equations (\ref{condition1}) and
(\ref{condition2}) imply that only two components of spin-tensor are independent, as it should be for an elementary spin one-half particle.

%

We could use Poisson brackets to obtain the Hamiltonian equations, $\dot z=\{z , H\}$, for the variables of physical
sector $z=(x, P, S)$. But in this case we are forced to work with rather inconvenient Hamiltonian (\ref{Hamiltonian}).
Instead, we construct the Dirac bracket associated with second-class constraints $T_3$ and $T_4$.  It is convenient to
denote $\{T_3, T_4\}=-\frac{1}{8\triangle}$, where $\triangle=\frac{-2}{16P^2-(\theta S)}$, then $\triangle\approx a$
on the surface of mass-shell constraint $T_1=0$. The Dirac bracket reads
\begin{eqnarray}\label{DB}
&\{A , B \}_D = \nonumber \\
&\{A , B\} -8\triangle\left[ \{ A , T_3\} \{T_4 , B\} - \{ A , T_4\}\{ T_3 , B\} \right] \,.
\end{eqnarray}
By construction, the Dirac bracket of any variable with the constraints vanishes, so $T_3$ and $T_4$ can be omitted
from the Hamiltonian. The first-class constraints $T_2$ and $T_5$ can be omitted as well, since brackets of the
variables $x, P$ and $S$ with them vanish on the constraint surface. In the result we arrive at a simple Hamiltonian
\begin{eqnarray}\label{Hamiltonian.0}
H_0= \frac{\lambda_1}{2} \left( P^2 + m^2c^2\right) \,,
\end{eqnarray}
which looks like that of a free point particle. All the information on spin and interaction is encoded now in the Dirac
bracket. In particular, equations of motion are obtained according the rule $\dot z=\{z, H_0\}_D$.

Poisson brackets of our variables with $T_3$ and $T_4$ are
\begin{eqnarray}\label{PB.A.3}
&\{ x^\mu , T_3 \} = \omega^\mu , \quad
 \{x^\mu , T_4 \} = \pi^\mu ,   \cr
&\{ P_\alpha , T_3 \} = -\frac 14 \theta_{\alpha\beta} \omega^\beta + \Gamma^\lambda_{\alpha\beta} P_\lambda\omega^\beta , \cr
&\{ P_\alpha , T_4 \} = - \frac 14 \theta_{\alpha\beta} \pi^\beta + \Gamma^\lambda_{\alpha\beta} P_\lambda\pi^\beta ,   \cr
&\{ S^{\mu\nu} , T_3 \} = 2 P^{[\mu} \omega^{\nu]} +  \Gamma^{[\mu}_{\alpha\beta} S^{\nu]\alpha} \omega^\beta , \cr
&\{ S^{\mu\nu} , T_4 \} = 2 P^{[\mu} \pi^{\nu]} +  \Gamma^{[\mu}_{\alpha\beta} S^{\nu]\alpha} \pi^\beta \, .
\end{eqnarray}
Using these expressions in (\ref{DB}), we obtain manifest form of the Dirac brackets
\begin{eqnarray}\label{DB.1}
&&\{ x^\mu , x^\nu\}_D = 4\triangle S^{\mu \nu}  ,  \nonumber  \\
&&\{ P_\mu , P_\nu \}_D =  -\frac 14 \theta_{\mu\nu} + 4\triangle (\Gamma P)_{\mu\alpha}  S^{\alpha\beta}  (\Gamma
P)_{\beta\nu}  \cr
&&-\frac{\triangle}{8} \left(  \theta_{\mu\alpha}  S^{\alpha\beta}\left[\theta_{\beta\nu} +4(\Gamma P)_{\beta\nu} \right]  -(\mu\leftrightarrow\nu) \right) , \nonumber\\
&&\{ x^\mu , P_\alpha \}_D=  \delta^\mu_\alpha + \triangle S^{\mu\beta}\left[\theta_{\beta\alpha} +4(\Gamma P)_{\beta\alpha} \right] \,,  \nonumber\\
&&\{x^\mu , S^{\alpha \beta} \}_D = -8\triangle \left[ S^{\mu[\alpha} P^{\beta]} -\frac 12 S^{\mu\sigma} \Gamma^{[\alpha}_{\sigma\lambda} S^{\beta]\lambda} \right]  \, , \nonumber \\
&&\{P_\alpha , S^{\mu\nu}\}_D = - \Gamma^{[\mu}_{\alpha\sigma} S^{\nu]\sigma}  \nonumber \\
&&+ \triangle\left[\theta_{\alpha\beta} +4(\Gamma P)_{\alpha\beta} \right] \left(2S^{\beta[\mu} P^{\nu]}-S^{\beta\eta} \Gamma^{[\mu}_{\eta \lambda} S^{\nu]\lambda}\right)
 \,, \nonumber \\
&&\{ S^{\mu\nu} , S^{\alpha\beta} \}_D= \{ S^{\mu\nu} , S^{\alpha\beta}  \} \\
&&-8\triangle\left[ 2\left( P^{\mu}P^{\alpha} S^{\beta\nu} -P^{\mu}P^{\beta} S^{\alpha\nu}-P^{\nu}P^{\alpha} S^{\beta\mu}+P^{\nu}P^{\beta} S^{\alpha\mu} \right)  \right. \nonumber \\
&&\left. - P^{ [\mu } S^{ \nu] \lambda} \Gamma^{ [\alpha }_{ \lambda\sigma } S^{ \beta]\sigma} + P^{ [\alpha } S^{ \beta] \lambda} \Gamma^{ [\mu }_{ \lambda\sigma } S^{ \nu]\sigma}  -\frac 12 \Gamma_{\sigma\lambda}^{[\mu} S^{\nu] \sigma} S^{\lambda\rho}\Gamma_{\rho\epsilon}^{[\alpha} S^{\beta]\epsilon} \right]\,. \nonumber
\end{eqnarray}
Their right hand sides do not contain explicitly the variables $\omega$ and $\pi$, so the brackets form a closed
algebra for the set $(x, P, S)$.

The Dirac brackets remain different from the Poisson brackets even in the limit of a free theory,
$g_{\mu\nu}\rightarrow\eta_{\mu\nu}$. In particular, in the sector of canonical variables $x$ and $p$ we have
\begin{eqnarray}\label{freeb}
&\{x^\mu, x^\nu\}_D=-\frac{S^{\mu\nu}}{2p^2}, \nonumber \\
&\{x^\mu, p^\nu\}_D=\eta^{\mu\nu}, \quad \{p^\mu, p^\nu\}_D=0.
\end{eqnarray}
Hence, account of spin leads to deformation of the phase-space symplectic structure: the position variables of
relativistic spinning particle obey the noncommutative bracket, with the noncommutativity parameter being proportional
to the spin-tensor. This must be taken into account in construction of quantum mechanics of a spinning particle
\cite{DPM2, LHuang1706}. In particular, for an electron in  electromagnetic field, the spin-induced noncommutativity
explains the famous one-half factor in the Pauli equation without appeal to the Thomas precession, Dirac equation or to
the Foldy-Wouthuysen transformation, see \cite{DPM2016}. Besides, for a spinning body in gravitational field, the
spin-induced noncommutativity clarifies the discrepancy in expressions for three-acceleration obtained by different
methods, see \cite{DPW2}.

Using the Dirac brackets together with the Hamiltonian (\ref{Hamiltonian.0}), we obtain equations of motion
\begin{eqnarray}\label{motion-x-DB}
&&\dot x^\mu=\{x^\mu , H_0 \}_D= \lambda_1\left[ P^\mu +aS^{\mu\beta}\theta_{\beta\alpha}P^\alpha\right] \, ,\cr 
&&\dot P_\mu =\{ P_\mu , H_0 \}_D = \left(-\frac 14 \theta_{\mu\nu}+(\Gamma P)_{\mu\nu}\right)\lambda_1\left[ P^\nu
+aS^{\nu\beta}\theta_{\beta\alpha}P^\alpha\right] \cr 
&&\qquad = -\frac 14 \theta_{\mu\nu} \dot x^\nu+\Gamma^\alpha_{\mu\nu}
P_\alpha \dot x^\nu \, ,  \cr 
&&\dot S^{\mu\nu}= \{S^{\mu\nu} , H_0\}_D \nonumber \\
&&\qquad = \left(2P^\mu\delta^\nu{}_\alpha
-\Gamma^\mu_{\alpha\sigma} S^{\sigma \nu}\right)\lambda_1\left[ P^\alpha
+aS^{\alpha\beta}\theta_{\beta\gamma}P^\gamma\right]-(\mu\leftrightarrow\nu)  \nonumber \\
&&\qquad = 2P^{[\mu} \dot x^{\nu]}  -\Gamma^\mu_{\alpha\sigma} S^{\sigma \nu} \dot x^\alpha   -
\Gamma^\nu_{\alpha\sigma}S^{\mu\sigma} \dot x^\alpha  \, .
\end{eqnarray}
They can be rewritten in a manifestly general-covariant form as follows:
\begin{eqnarray}
\dot x^\mu &=& \lambda_1\left(\delta^\mu{}_\nu +aS^{\mu\beta}\theta_{\beta\nu}\right)P^\nu \, , \label{motion-x.1} \\
\nabla P_\mu &=&- \frac{1}{4} R_{\mu\nu\alpha\beta} S^{\alpha\beta}\dot  x^\nu\equiv-\frac{1}{4}\theta_{\mu\nu}\dot
x^\nu \, ,  \label{motion-P-2} \\
\nabla S^{\mu\nu} &=& 2(P^\mu \dot x^\nu - P^\nu \dot x^\mu )\, . \label{motion-S}
\end{eqnarray}
Some relevant comments are in order. \par \noindent 1. {\it Comparison with MPTD equations.} Despite the fact that the
vector model has been initially constructed as a theory of an elementary particle of spin one-half, it turns out to be
suitable to describe a rotating body in general relativity in the pole-dipole approximation \cite{Trautman2002,
Dixon1964}. Indeed, the equations (\ref{motion-P-2}) and (\ref{motion-S}) coincide with Dixon equations of the body
(our spin is twice of that of Dixon), while our constraint (\ref{condition1}) is just the Tulczyjew spin supplementary
condition\footnote{While the variational problem dictates \cite{hanson1974} the equation (\ref{condition1}), in the
multipole approach there is a freedom in the choice of a spin supplementary condition, related with the freedom in the
choice of a representative point $x^\mu$ describing position of the body \cite{Papapetrou:1951pa, Tulc, pirani:1956}.
Different conditions lead to the same results in $\frac{1}{c^2}$\,-approximation, see \cite{Schiff1960.1, Dixon1964,
Connell1975}. }. Besides, the Hamiltonian equation (\ref{motion-x.1}) can be identified with the velocity-momentum
relation, implied by MPTD-equations, see \cite{DWGR2016} for a detailed comparison. The only difference is that values
of momentum and spin are conserved quantities of MPTD equations, while in the vector model they are fixed by
constraints. In summary \cite{DWGR2016}, to study the class of trajectories of a body with $\sqrt{-P^2}=k$ and
$S^2=\beta$, we can use our spinning particle with $m=\frac{ k}{c}$ and $\alpha=\frac{\beta}{8}$.  \par

\noindent 2. {\it Ultra relativistic limit.} Using the Landau-Lifshitz $1+3$\,-decomposition \cite{bib16},  we showed
in \cite{DWGR2016} that MPTD equations yield a paradoxical behavior in ultra relativistic limit: three-dimensional
acceleration of the particle grows with its speed, and diverges as $|{\bf v}|\rightarrow c$. In the next section, we
improve this by adding a non-minimal spin-gravity interaction through the gravimagnetic moment. \par

\noindent 3. {\it Analogy between gravitation and electromagnetism.} Many people mentioned remarkable analogies between
gravitation and electromagnetism in various circumstances \cite{Thirring1918.1, Wald1972, Pomeransky1998, Costa:2012cy, Natario:2007pu}. Here we
observe an analogy, comparing (\ref{motion-x.1})-(\ref{motion-S}) with equations of motion of spinning particle (with
null gyromagnetic ratio) \cite{deriglazov2014Monster} in electromagnetic field with the strength $F_{\mu\nu}$
\begin{eqnarray}
&&\dot x^\mu = \lambda_1\left(\delta^\mu{}_\nu +aS^{\mu\beta}F_{\beta\nu}\right)P^\nu \, , \cr
&& \mbox{where} \quad a=\frac{-2e}{4m^2c^3-e(SF)}, \label{x.1} \\
&&\dot P_\mu =\frac{e}{c} F_{\mu\nu}\dot  x^\nu \, ,  \label{x.2} \\
&&\dot S^{\mu\nu} = 2P^{[\mu} \dot x^{\nu]} \, . \label{x.3}
\end{eqnarray}
One system just turns into another if we identify $\theta_{\mu\nu}\equiv R_{\mu\nu\alpha\beta}S^{\alpha\beta}\sim
F_{\mu\nu}$, and set $e=-\frac{c}{4}$. That is a curvature influences trajectory of a spinning particle in the same way
as an electromagnetic field with the strength $\theta_{\mu\nu}$. We now use this analogy to construct a non-minimal
spin-gravity interaction.

\section{Rotating body with gravimagnetic moment}\label{sect.3}

The Hamiltonian (\ref{Hamiltonian}) is a combination of constraints, so the Hamiltonian formulation of our model is
completely determined by the set of constraints (\ref{primary}), (\ref{primaryb}), and by the expression
(\ref{canonical-P}) for canonical momentum $P^\mu$ through the conjugated momentum $p^\mu$. We observe that algebraic
properties of the constraints do not change, if we replace the mass-shell constraint $T_1=P^2+m^2c^2$ by $\tilde
T_1=P^2+f(x, P, S)+ m^2c^2$, where $f(x^\mu, P^\nu, S^{\mu\nu})$ is an arbitrary scalar function. Indeed, in the
modified theory $T_3$ and $T_4$ remain the second-class constraints, while $T_2$, $T_5$ and the combination
$\tilde T_1-\{T_3, T_4\}^{-1}\{\tilde T_1, T_4\}T_3+\{T_3, T_4\}^{-1}\{\tilde T_1, T_3\}T_4$,
form a set of first-class constraints. If we confine ourselves to the linear in curvature and quadratic in spin
approximation, the only scalar function $f$, which can be constructed from the quantities at our disposal is
$\frac{\kappa}{16} R_{\mu\nu\alpha\beta}S^{\mu\nu}S^{\alpha\beta}\equiv\kappa
R_{\mu\nu\alpha\beta}\omega^\mu\pi^\nu\omega^\alpha\pi^\beta$, where $\kappa$ is a dimensionless parameter. The
resulting constraint
\begin{equation}\label{Hamiltonian-k}
\tilde T_1= P^2 + \frac{\kappa}{16}(\theta S) + m^2c^2=0 \, ,
\end{equation}
is similar to the Hamiltonian $\frac{\lambda_1}{2} \left( P^2-\frac{e g}{c}(FS) + m^2c^2\right)$ of a spinning particle
interacting with electromagnetic field through the gyromagnetic ratio $g$, see \cite{deriglazov2014Monster}. In view of
this similarity, the interaction constant $\kappa$ is called gravimagnetic moment \cite{Khriplovich1989,
Pomeransky1998}, and we expect that non-minimally interacting theory with the Hamiltonian (\ref{Hamiltonian-k}) could
be consistent generalization of MPTD equations. The consistency has been confirmed in \cite{DWGR2016}, where we
presented the Lagrangian action of a spinning particle that implies the constraints (\ref{Hamiltonian-k}) and
(\ref{primaryb}) in Hamiltonian formalism.

Poisson brackets of the constraints $\tilde T_1$, $T_3$ and $T_4$ read
\begin{eqnarray}
\{\bar T_1, T_3\} &&= \frac 12(1-\kappa)(\omega\theta P) \cr
&&+\kappa \omega^\sigma(\nabla_\sigma R_{\mu\nu\alpha\beta}) \omega^\mu\pi^\nu\omega^\alpha\pi^\beta  \, . \label{mult-7} \\
\{\bar T_1 , T_4 \} && =\frac 12  (1-\kappa)(\pi\theta P) \cr
&&+\kappa \pi^\sigma(\nabla_\sigma R_{\mu\nu\alpha\beta}) \omega^\mu\pi^\nu\omega^\alpha\pi^\beta  \, . \label{mult-6} \\
\{T_3, T_4\}&&= P^2 - \frac{1}{16}(\theta S) \approx -8\bar a, \cr
&& \mbox{where} ~ \bar a=\frac{2}{16m^2c^2+(\kappa+1)(\theta S)}.
\label{mult-11}
\end{eqnarray}
These expressions must be substituted in place of terms $\frac 12 (\omega\theta P)$, $\frac 12 (\omega\theta P)$ and
$a$ in the Table \ref{algebra-constraints}. The Dirac brackets (\ref{DB.1}), being constructed with help of $T_3$ and
$T_4$, remain valid in the modified theory. Our new Hamiltonian is $H=\frac{\lambda}{2}H_0+
\frac{\lambda}{2}H_\kappa$, with $H_0$ from (\ref{Hamiltonian.0}) and $H_\kappa=\frac{\kappa}{16}(\theta S)$. Hence, to
obtain the manifest form of equations of motion $\dot z=\{z, H_0\}_D+\{z, H_\kappa\}_D$, we only need to compute the
brackets $\{z, H_\kappa\}_D$. They are
\begin{eqnarray}
\{ x^\mu , H_\kappa \}_D =&& -\lambda_1\kappa \bar a \left[ S^{\mu\alpha}\theta_{\alpha\beta} P^\beta \right. \cr
&& \left. - \frac {1}{8} S^{\mu\nu} (\nabla_\nu R_{\alpha\beta\sigma\lambda})S^{\alpha\beta} S^{\sigma\lambda}  \right] \, ,\\ \label{Hk-x}
\{ P_\mu , H_\kappa \}_D=  &&-\frac 14 \theta_{\mu\alpha} \{x^\alpha , H_\kappa \} _D+ \Gamma_{\mu\alpha}^\beta P_\beta  \{ x^\alpha , H_\kappa\}_D \cr
&&- \frac{\lambda_1 \kappa}{32} (\nabla_\mu R_{\alpha\beta\sigma\lambda}) S^{\alpha\beta}S^{\sigma\lambda} \, ,\label{Hk-P} \\
\{ S^{\mu\nu}, H_\kappa \}_D = && \frac{ \kappa \lambda_1}{4} \theta^{[\mu}_{\  \alpha} S^{\nu]\alpha} +  2P^{[\mu} \{x^{\nu]}, H_\kappa \}_D \cr
&&- \left( \Gamma^\mu_{\alpha\beta} S^{\alpha\nu} + \Gamma^\nu_{\alpha\beta}S^{\mu\alpha} \right) \{ x^\beta , H_\kappa \}_D  \, .\label{Hk-S}
\end{eqnarray}
Adding them to the equations $\dot z=\{z, H_0\}_D$ given in  (\ref{motion-x.1})-(\ref{motion-S}), we arrive at the dynamical equations
\begin{eqnarray}
\dot x^\mu&&= \lambda_1  \left[\delta^\mu{}_\nu -\bar a(\kappa -1) S^{\mu\alpha}\theta_{\alpha\nu}\right] P^\nu  \nonumber \\
&& + \frac{\lambda_1 \kappa \bar a}{8} S^{\mu\nu}(\nabla_\nu  R_{\alpha\beta\sigma\lambda}) S^{\alpha\beta}S^{\sigma\lambda} \, , \label{motion-x-k}  \\
\nabla P_\mu &&= -\frac 14 \theta_{\mu\nu}\dot x^\nu - \frac{\lambda_1\kappa}{32} (\nabla_\mu  R_{\alpha\beta\sigma\lambda}) S^{\alpha\beta}S^{\sigma\lambda} \, , \label{motion-P-k} \\
\nabla S^{\mu\nu}&& = 2P^{[\mu} \dot x^{\nu]} +\frac{\lambda_1\kappa}{4}\theta^{[\mu}{}_{\alpha} S^{\nu]\alpha}  \, . \label{motion-S-k}
\end{eqnarray}
Together with the constraints (\ref{condition1}), (\ref{condition2}), and (\ref{Hamiltonian-k}), they give complete
system of Hamiltonian equations of spinning particle with gravimagnetic moment $\kappa$. As it should be, our equations
reduce to MPTD equations (\ref{motion-x.1})-(\ref{motion-S}) when $\kappa=0$. Comparing the two systems, we see that
the non-minimal interaction yields quadratic and cubic in spin corrections to MPTD equations.

The equations (\ref{motion-x-k})-(\ref{motion-S-k}) are greatly simplified for a particle with unit gravimagnetic
moment, $\kappa=1$ (gravimagnetic particle). It has a qualitatively different behavior as compared with MPTD particle.
First,  gravimagnetic particle has an expected behavior in the ultra relativistic limit \cite{DWGR2017,DWGR2016}:
three-dimensional acceleration of the particle and angular velocity of precession remain finite as $|{\bf
v}|\rightarrow c$, while the longitudinal acceleration vanishes in the limit. Second, at low velocities, taking
$\kappa=1$ and keeping only the terms which may give a contribution in the leading post-Newton approximation,
$\sim\frac{1}{c^2}$, we obtain from (\ref{motion-x-k})-(\ref{motion-S-k}) the approximate equations
\begin{eqnarray}\label{xk.1}
&\dot x^\mu= \lambda_1 P^\mu \, ,  ~ \nabla P_\mu = -\frac 14 \theta_{\mu\nu}\dot x^\nu -\frac{\lambda_1}{32}
(\nabla_\mu  R_{\alpha\beta\sigma\lambda}) S^{\alpha\beta}S^{\sigma\lambda}  \, , \nonumber \\
&\nabla S^{\mu\nu}=\frac{\lambda_1}{4}\theta^{[\mu}{}_{\alpha} S^{\nu]\alpha}  \,,
\end{eqnarray}
while MPTD equations ($\kappa=0$) in the same approximation read
\begin{eqnarray}\label{x.111}
\dot x^\mu = \lambda_1 P^\mu \, ,  \quad
\nabla P_\mu =-\frac{1}{4}\theta_{\mu\nu}\dot
x^\nu \, ,  \quad
\nabla S^{\mu\nu} = 0 \, .
\end{eqnarray}
In Sect. \ref{sect.4}, we compute $\frac{1}{c^2}$ corrections due to the extra-terms appeared in (\ref{xk.1}).

\noindent{\bf Conserved charges.}
%
In curved space which possesses some isometry, MPTD equations admit a  constant of motion (see, for example, \cite{DW2015.1})
\begin{equation}\label{constant-motion}
J^{(\xi)}=P^\mu\xi_\mu - \frac{1}{4}S^{\mu\nu}\nabla_\nu\xi_\mu \, ,
\end{equation}
where $\xi_\mu$ is Killing vector which generates the isometry, i.e.,  $\nabla_\mu\xi_\nu +\nabla_\nu\xi_\mu=0$. Let us
show that $J^{(\xi)}$ remains a constant of motion when the gravimagnetic interaction is included. Using
(\ref{motion-P-k}) and (\ref{motion-S-k}), we obtain by direct calculation
\begin{equation}\label{charge2}
\dot J^{(\xi)}= \frac{\kappa \lambda_1}{8} \left[ S^{\alpha\beta}R^\mu_{\ \sigma\alpha\beta} S^{\sigma\nu}\nabla_\nu
\xi_\mu - \frac 14 S^{\alpha\beta}S^{\sigma\lambda}\xi^\mu \nabla_\mu R_{\alpha\beta\sigma\lambda} \right] \, .
\end{equation}
Using the  Bianchi identities we find the relation
\begin{equation}\label{charge3}
S^{\alpha\beta}S^{\sigma\lambda}\xi^\mu \nabla_\mu  R_{\alpha\beta\sigma\lambda} =2 S^{\alpha\beta} S^{\sigma\nu}
\xi^\mu \nabla_\sigma R_{\mu\nu\alpha\beta} \, .
\end{equation}
Derivative of a curvature tensor is related with derivative of a Killing vector by the formula
$\xi^\mu \nabla_\sigma R_{\alpha\beta\nu\mu} -  \xi^\mu \nabla_\nu R_{\alpha\beta\sigma\mu}= R_{\alpha \beta \sigma}^{\
\  \  \  \mu} \nabla_\nu \xi_\mu - R_{\alpha \beta \nu}^{\ \ \ \ \mu} \nabla_\sigma \xi_\mu  + R_{\sigma\nu\alpha}^{\ \
\ \ \mu} \nabla_\beta \xi_\mu  - R_{\sigma \nu \beta}^{\ \ \ \ \mu} \nabla_\alpha \xi_\mu$.
Contracting twice with the spin tensor  we obtain
\begin{equation}\label{charge5}
S^{\alpha\beta} S^{\sigma\nu} \xi^\mu \nabla_\sigma R_{\mu\nu\alpha\beta} =2 S^{\alpha\beta} R^{\mu}_{\
\sigma\alpha\beta} S^{\sigma\nu} \nabla_\nu\xi_\mu.
\end{equation}
Using this expression in  (\ref{charge3}), we obtain $S^{\alpha\beta}S^{\sigma\lambda}\xi^\mu \nabla_\mu
R_{\alpha\beta\sigma\lambda}=4S^{\alpha\beta} R^{\mu}_{\ \sigma\alpha\beta} S^{\sigma\nu} \nabla_\nu\xi_\mu$. This
implies that the right hand side of (\ref{charge2}) vanishes, so $\dot J^{(\xi)}=0$. Thus,  the quantity
(\ref{constant-motion}) represents a constant of motion  of a spinning particle with gravimagnetic moment.

\noindent{\bf Lagrangian System of equations of motion.}
%
Since we are interested in the influence of non-minimal spin-gravity interaction on trajectory and spin of the
particle, we eliminate the momenta $P^\mu$ and the auxiliary variable $\lambda_1$ from the equations
(\ref{motion-x-k})-(\ref{motion-S-k}), obtaining their Lagrangian form.  In the equation (\ref{motion-x-k}), which
relates velocity and momentum, appeared the matrix
\begin{equation}\label{gmm11}
T^{\alpha}{}_{\nu} \equiv \delta^\alpha{}_\nu -(\kappa-1)\bar aS^{\alpha \sigma}\theta_{\sigma\nu} \, .
\end{equation}
Using the identity
$(S\theta S)^{\mu\nu}=-\frac{1}{2} (S^{\alpha\beta}\theta_{\alpha\beta})S^{\mu\nu}$,
we find inverse\footnote{We point out that the analogous matrix present in MPTD equations can not be explicitly
inverted in the multipole approach.} of the matrix $T$
\begin{eqnarray}\label{gmm14}
\tilde T^{\alpha}{}_{\nu} \equiv \delta^\alpha{}_\nu+(\kappa-1)bS^{\alpha \sigma}\theta_{\sigma\nu} \, ,
~ b=\frac{1}{8m^2c^2+\kappa(S\theta)} \,.
\end{eqnarray}
Using (\ref{gmm14}), we solve (\ref{motion-x-k}) with respect to $P^\mu$. Using the resulting expression in the
constraint (\ref{Hamiltonian-k}), we obtain
$\lambda_1=\frac{\sqrt{-\dot x G\dot x}}{m_r c}$,
where $m_r^2\equiv m^2 + \frac{\kappa}{16 c^2} (S\theta) -\kappa^2 Z^2$ is the radiation mass in gravitational field.
By $Z^\mu$ we have denoted the vector, which vanishes in spaces with covariantly-constant curvature,
$Z^\mu=\frac{b}{8c}S^{\mu\sigma}(\nabla_\sigma R_{\alpha\beta\rho\delta})S^{\alpha\beta}S^{\rho\delta}$. Besides, in
the expression for $\lambda_1$ appeared a kind of effective metric $G$ induced by spin-gravity interaction along the
world-line,
$G_{\mu\nu} = \tilde T^\alpha{}_{\mu} g_{\alpha\beta} \tilde T^\beta{}_{\nu}$.
Only for the gravimagnetic particle ($\kappa=1$), the effective metric reduces to the original one. Using
(\ref{motion-x-k}) and  (\ref{gmm14}), we obtain expression for momentum in terms of velocity
\begin{eqnarray}\label{gmm17}
P^\mu = \frac{m_r c}{\sqrt{-\dot x G \dot x}} \tilde T^{\mu}{}_{\nu} \dot x^\nu-\kappa c Z^\mu.
\end{eqnarray}
We substitute this $P^\mu$ into (\ref{motion-P-k}) and (\ref{motion-S-k}), arriving at the Lagrangian equations of our
spinning particle with gravimagnetic moment $\kappa$
\begin{eqnarray}
\nabla\left[ \frac{m_r }{\sqrt{-\dot x G \dot x}} \tilde T^\mu{}_\nu \dot x^\nu \right] =
-\frac{1}{4c} \theta^\mu{}_\nu\dot x^\nu  -  \kappa\frac{\sqrt{-\dot x G\dot x}}{32m_r c^2}
\nabla^\mu(S\theta) \nonumber \\ 
+\kappa\nabla Z^\mu, \qquad \label{gml1} \\
\nabla S^{\mu\nu} =- \frac{\kappa\sqrt{-\dot x G\dot x}}{4m_r c }(\theta S)^{[\mu\nu]} -\frac{2m_r c
(\kappa-1)b}{\sqrt{-\dot x
G\dot x}}\dot x^{[\mu} (S\theta\dot x)^{\nu]}  \nonumber \\ 
+ 2\kappa c\dot x^{[\mu}Z^{\nu]}. \qquad \label{gml2}
\end{eqnarray}

\section{Leading post-Newtonian corrections due to unit gravimagnetic moment}\label{sect.4}
Taking $\kappa=1$ in (\ref{gml1}) and (\ref{gml2}), we obtain equations of our gravimagnetic body
\begin{eqnarray}
\nabla \left[ \frac{m_r\dot x^\mu}{\sqrt{-\dot x g \dot x}} \right] && = -\frac{1}{4c} \theta ^\mu{}_{\nu}\dot x^\nu  -
\frac{\sqrt{-\dot x g \dot x}}{32m_r c^2}
\nabla^\mu(S\theta) \nonumber \\
&&+\nabla Z^\mu\, , \label{xk1}\\
\nabla S^{\mu\nu} &&=-\frac{\sqrt{-\dot x g \dot x}}{4m_r c}(\theta S)^{[\mu\nu]} +2 c\dot x^{[\mu}Z^{\nu]} \, . \label{Sk1}
\end{eqnarray}
To test these equations, we compute the leading relativistic corrections due to unit gravimagnetic moment to the
trajectory and precession of a gyroscope,  orbiting around a rotating spherical body of mass $M$ and angular momentum
${\bf J}$. To this aim, we write equations of motion implied by (\ref{xk1}) and (\ref{Sk1}) for the three-dimensional
position $x^i(t)$ and for the spin-vector
\begin{eqnarray}\label{d3s}
&{\bf S}=\frac12\left( S^{23}, S^{31}, S^{12}\right), \quad \mbox{or} \quad
S_i(t)=\frac14\epsilon_{ijk}S^{jk}(t), \cr
& S^{ij}=2\epsilon^{ijk}S_k,
\end{eqnarray}
as functions of the coordinate time $t=\frac{x^ 0}{c}$. Due to the  reparametrization invariance, the desired equations are obtained by
setting $\tau=t$ in (\ref{xk1}) and (\ref{Sk1}). We consider separately the trajectory and the spin.

\noindent{\bf Trajectory.}
We denote
$v^\mu\equiv \frac{dx^\mu}{dt}=(c, {\bf v})$,
so
$\sqrt{-\dot x g \dot x}=\sqrt{-v g v}=\sqrt{-c^2g_{00}-2cg_{0i}v^i-g_{ij}v^iv^j}$.  
The temporal and spatial parts of Eq. (\ref{xk1}) read
\begin{eqnarray}\label{Ap000}
&&\frac{d}{dt}\left[ \frac{m_r}{\sqrt{-vgv}} \right] +\frac{m_r}{c\sqrt{-vgv}}\Gamma^0{}_{\mu\nu}v^\mu v^\nu \nonumber \\
&&\qquad = -\frac{1}{4c^2} \theta ^0{}_{\nu}v^\nu  
- \frac{\sqrt{-v g v}}{32m_r c^3} \nabla^0(S\theta)+\frac{1}{c}\nabla_t Z^0 \, , \nonumber\\
 &&\frac{d^2 x^i}{dt^2} + \Gamma^i_{\mu\nu} v^\mu v^\nu  + \frac{v^i\sqrt{-vgv}}{m_r}  \frac{d}{dt}\left[ \frac{m_r}{\sqrt{-vgv}} \right] \nonumber \\
&&\qquad =-\frac{{\sqrt{-v g v}}}{4m_r c} \theta ^i{}_{\nu}v^\nu +\frac{v g v}{32m^2_r c^2}\nabla^i(S\theta)
+\frac{\sqrt{-vgv}}{m_r}\nabla_t Z^i \, . \nonumber
\end{eqnarray}
Using the first equation in the second one, we avoid the necessity to compute time derivative in the second term, and obtain
\begin{eqnarray}\label{Ap1}
\frac{d^2 x^i}{d t^2}&&=-\Gamma^i{}_{\mu\nu}v^\mu v^\nu+\frac{v^i}{c}\Gamma^0{}_{\mu\nu}v^\mu v^\nu  \cr 
&& - \frac{\sqrt{-v gv}}{4m_rc}\left[\theta ^i{}_{\nu}v^\nu-\frac{v^i}{c}\theta ^0{}_{\nu}v^\nu \right]  \cr 
&&+ \frac{v g v}{32m^2_r c^2}\left[\nabla^i(S\theta)-\frac{v^i}{c}\nabla^0(S\theta) \right] \cr
&& +\frac{\sqrt{-v g v}}{m_r}\left[\nabla_t
Z^i -\frac{v^i}{c}\nabla_t Z^0\right]\, .  \qquad \quad
\end{eqnarray}
Now we assume a non relativistic motion, $\frac{v}{c}<< 1$, and expand all quantities in (\ref{Ap1}) in series with
respect to $\frac{1}{c}$. Typical metric of stationary spaces has the series of the form \cite{Weinberg}
\begin{eqnarray}
g_{00} = -1 + \ ^{2}g_{00} + \ ^{4} g_{00} + \ldots \nonumber \\
g_{ij} = \delta_{ij} + \ ^{2}g_{ij} + \ ^{4}g_{ij} + \ldots \\ \label{PNmetric2}
g_{i0} = \ ^{3}g_{i0} + \ ^{5} g_{i0} +\ldots \, , \nonumber
\end{eqnarray}
where $^{n}g_{\mu\nu}$ denotes the  term in $g_{\mu\nu}$ of order $ 1/c^n$.  As a consequence,  the series of
connection, curvature and its covariant derivative starts from $\frac{1}{c^2}$ or from higher order. In some details,
we have
\begin{eqnarray}
&&\Gamma^\mu_{\nu\alpha} = \ ^2\Gamma^\mu_{\nu\alpha} + \ ^4\Gamma^\mu_{\nu\alpha} + \ldots ~ \textrm{for} ~ \Gamma^i_{00}, ~ \Gamma^i_{mn}, ~ \Gamma^0_{0m} \, ,  \label{PN2} \\
&&\Gamma^\mu_{\nu\alpha} = \ ^3\Gamma^\mu_{\nu\alpha} +\ ^5\Gamma^\mu_{\nu\alpha} + \ldots ~   \textrm{for} ~ \Gamma^i_{0m}, ~ \Gamma^0_{00}, ~ \Gamma^0_{mn} \, , \label{PN2a}
\end{eqnarray}
\begin{eqnarray}
&&R^\mu_{\ \nu\alpha\beta} = \label{PN-R1} \\
&& \ ^2R^\mu_{\ \nu\alpha\beta} + \ ^4R^\mu_{\ \nu\alpha\beta} + \ldots ~  \textrm{for} ~ R^0_{\ mn0}, ~ R^0_{\ 0mn}, ~ R^i_{\ 0m0}, ~ R^i_{\ jmn} \, , \nonumber\\
&&R^\mu_{\ \nu\alpha\beta} = \label{PN-R2} \\
&& \ ^3R^\mu_{\ \nu\alpha\beta} + \ ^5R^\mu_{\ \nu\alpha\beta} + \ldots \quad   \textrm{for} ~ R^0_{\ imn}, ~ R^0_{\ 0m0}, ~ R^i_{\ 0mn}, ~ R^i_{\ jm0} \, .  \nonumber
\end{eqnarray}
Besides, for various quantities which appear in equations (\ref{xk1}) and  (\ref{Sk1}), we have the estimations
\begin{eqnarray}\label{Ap4}
&&\sqrt{-vgv}\sim c+\frac1c+\ldots \,, \qquad -vgv\sim c^2+1+\frac{1}{c^2}+\ldots \,,  \cr
&&m_r^2\sim m^2+\frac{1}{c^4}+\ldots, \qquad \theta_{\mu\nu}\sim\frac{1}{c^2}+\ldots, \cr
&&b\sim\frac{1}{c^2}+\ldots, \qquad Z^\mu \sim \frac{1}{c^5}+\ldots\,.
\end{eqnarray}
At last, the spin supplementary condition implies
\begin{eqnarray}\label{Ap4.1}
S^{i0}=\frac{1}{c}S^{ij}v^j+\ldots\,.
\end{eqnarray}
Keeping only the terms which may contribute up to order $\frac{1}{c^2}$ in the equation (\ref{Ap1}), we obtain
\begin{eqnarray}\label{Ap1.6}
&&\frac{d^2 x^i}{d t^2}=-\Gamma^i{}_{\mu\nu}v^\mu v^\nu+\frac{v^i}{c}\Gamma^0{}_{\mu\nu}v^\mu v^\nu  \nonumber \\
&& \quad + \frac{1}{4m} \left[v^i\theta ^0{}_{0}-c\theta^i{}_0-\theta^i{}_jv^j\right]- \frac{1}{32m^2}\nabla^i(S\theta)\, .
\end{eqnarray}
The terms on right-hand side of this equation are conveniently grouped according to their origin
\begin{eqnarray}\label{PN-F1}
\frac{d^2 \mathbf{x}}{dt^2}= \mathbf{a}_{\Gamma} + \mathbf{a}_{R} + \mathbf{a}_{\nabla R} \, .
\end{eqnarray}
Here $\mathbf{a}_\Gamma$ is the contribution due to connection, $\mathbf{a}_{R}$  comes from interaction between spin
and space-time curvature, and  $\mathbf{a}_{\nabla R}$ is the contribution which involves derivatives of the Riemann
tensor. Using (\ref{PN2})-(\ref{PN-R2}) we obtain
\begin{eqnarray}\label{ka1}
a^i_\Gamma \equiv&& -\Gamma^i_{\alpha\beta} v^\alpha v^\beta + \frac{v^i}{c}\Gamma^0_{\alpha\beta}v^\alpha v^\beta \nonumber \\
=&& -c^2 \ ^2\Gamma^i_{00}-\ ^2\Gamma^i_{mn}v^n v^m  + 2v^i\  ^2\Gamma^0_{m0} v^m \nonumber \\
&&  -c^2 \ ^4\Gamma^i_{00} + cv^i \ ^3\Gamma^0_{00} -  2c \ ^3\Gamma^i_{m0}v^m,
\end{eqnarray}
\begin{eqnarray}\label{ka2}
a^i_{R} \equiv&&  \frac{1}{4m} \left[v^i\theta ^0{}_{0}-c\theta^i{}_0-\theta^i{}_jv^j\right]   \nonumber\\
=&& -\frac{1}{4m} \left[ 2\ ^2R^i_{\ 0m0} S^{mn} v^n + \ ^2R^i_{\ kmn} S^{mn} v^k \right. \nonumber \\ 
&&\left.  -\ ^2R^0_{\ 0mn} S^{mn} v^i  \right]  -\frac{c}{4m} \ ^3R^i_{\ 0mn} S^{mn} \, ,
\end{eqnarray}
\begin{eqnarray}\label{ka3}
a^i_{\nabla R} \equiv&& -\frac{1}{32 m^2} g^{i\sigma}\nabla_\sigma R_{\alpha\beta\mu\nu}S^{\alpha\beta}
S^{\mu\nu} \nonumber \\
=&&-\frac{1}{32m^2} \partial_i \ ^2R_{jklm} S^{jk}S^{lm} \,.
\end{eqnarray}
As a concrete example of an external gravitational field, we take a stationary, asymptotically flat metric in the
post-Newtonian approximation up to order $\frac{1}{c^4}$ \cite{Weinberg}
\begin{eqnarray}\label{PN-metric}
ds^2 = &&\left(- 1 +\frac{2GM}{c^2r}-\frac{2G^2M^2}{c^4r^2} \right)(dx^0)^2 \nonumber \\
&&- 4G\frac{\epsilon_{ijk}J^j x^k}{c^3r^3} dx^0 dx^i \nonumber \\
&&+\left( 1+ \frac{2GM}{c^2r} +\frac{3G^2M^2}{2c^4r^2}\right)dx^idx^i \,.
\end{eqnarray}
It can be obtained taking the asymptotic form of the Kerr metric for a large radial coordinate \cite{Straumann}.  With
this metric, the equations  (\ref{ka1})-(\ref{ka3}) are\footnote{The first two terms in ${\bf a}_R$ can be written also
as follows: $-3\frac{GM}{m c^2r^3} \left[ \left( \mathbf{v} \times \mathbf{S} \right) -2 \mathbf{\hat r} \left(
\mathbf{\hat r} \cdot\left( \mathbf{ v} \times \mathbf{ S} \right)\right) - \left( \mathbf{\hat r} \cdot \mathbf{v}
\right) \left( \mathbf{\hat r} \times \mathbf{S} \right)\right]$.}
\begin{eqnarray}\label{acc-4.0}
\mathbf{a}_{\Gamma} =&&  -\frac{MG}{r^2} \mathbf{\hat r} + \frac{4GM}{c^2r^2}(  \hat{\mathbf{r}}\cdot\mathbf{v} )
\mathbf{v} -\frac{GM}{c^2r^2} v^2\mathbf{\hat r} + \frac{4G^2M^2}{c^2r^3}\mathbf{\hat r} \nonumber \\
&&+  2 \frac {G}{c^2}\left[ \frac{3(\mathbf{J}\cdot \mathbf{\hat r}) \mathbf{\hat r} - \mathbf{ J}}{r^3}\right]  \times \mathbf{v} \, ,
\end{eqnarray}
\begin{eqnarray}\label{PN-F5}\label{acc-R.0}
\mathbf{a}_R  =&& 3\frac{GM}{mc^2r^3} \left[ (\hat{\bf r}\times{\bf v})(\hat{\bf r}\cdot{\bf S})+\hat{\bf r}({\bf
S}\cdot(\hat{\bf r}\times{\bf v}))\right] \nonumber \\
&& - \frac{1}{m} {\boldsymbol{\nabla}}  \left[  \frac {G}{c^2}\left(
\frac{3(\mathbf{J}\cdot \mathbf{\hat r}) \mathbf{\hat r} - \mathbf{ J}}{r^3}\right)  \cdot \mathbf{S} \right]  \, ,
\end{eqnarray}
\begin{equation}\label{acc-5.0}
\mathbf{a}_{\nabla R} = -\frac {1}{2m}  {\boldsymbol{\nabla}} \left[   \frac{G}{c^2} \left(\frac{M}{m}
\right)\left(\frac{3(\mathbf{ S} \cdot \mathbf{\hat r}) \mathbf{\hat r} - \mathbf{S} }{r^3}  \right) \cdot \mathbf{S}
\right] \, .
\end{equation}
We denoted by $\hat{\bf r}$ the unit vector in the direction of ${\bf r}$.

\par

\noindent {\bf Spin torque.} Setting $\kappa=1$ and $\tau=t\equiv\frac{x^0}{c}$ in the spatial part of Eq.
(\ref{Sk1}), this reads
\begin{eqnarray}
\frac{dS^{ij}}{dt}=&& -\Gamma^i_{\alpha\beta} v^\alpha S^{\beta j} - \Gamma^j_{\alpha\beta} v^\alpha S^{i \beta}
+\frac{\sqrt {-vgv}}{4m_r c} \theta^{[i}_{\ \alpha} S^{j]\alpha} \nonumber \\
&&+2c v^{[i} Z^{j]}.
\end{eqnarray}
For the spin-vector (\ref{d3s}), this equation implies
\begin{eqnarray}\label{PN-S-1}
\frac{dS^i}{dt} =&&-\frac12\epsilon^{ijk}\Gamma^j{}_{\mu\nu}v^\mu S^{\nu k}
-\frac{\sqrt{-v g v}}{8m_r c}\epsilon^{ijk}\theta^j{}_\nu S^{\nu k} \nonumber \\
&&+c\epsilon^{ijk}v^j Z^k \, .
\end{eqnarray}
Taking into account the equations (\ref{PNmetric2})-(\ref{Ap4}), we keep only the terms which may contribute up to order $\frac{1}{c^2}$
\begin{eqnarray}\label{PN-S-11}
\frac{dS^i}{dt} =&&-\frac12\epsilon^{ijk}\left[c\Gamma^j{}_{00}S^{0k}+v^n\Gamma^j{}_{nm}S^{mk}+c\Gamma^j{}_{0n}S^{nk}\right] \nonumber \\
&&-\frac{1}{8m} \epsilon^{ijk}\theta^j{}_n S^{nk} \nonumber\\
 =&&S^n \left( \ ^2\Gamma^n_{00} v^i  + \ ^2\Gamma^n_{ik} v^k \right) -
S^i \left( \ ^2\Gamma^k_{00} + \ ^2\Gamma^l_{kl} \right) v^k \nonumber \\
 && +c \ ^3\Gamma^k_{0i} S^k +
\frac{1}{2m} \epsilon_{mnl} \ ^2R^k{}_{imn}  S^kS^l \, .
\end{eqnarray}
The total torque on right hand side of this equation can be conveniently grouped as follows:
\begin{equation}\label{PN-S}
\frac{d\mathbf{S}}{dt} =  \boldsymbol{\tau}_v +  \boldsymbol{\tau}_J + \boldsymbol{\tau}_R \, ,
\end{equation}
where $\boldsymbol{\tau}_v$ contains the velocity-dependent terms, $\boldsymbol{\tau}_J$ depends on inner angular
momentum of central body, and $\boldsymbol{\tau}_R$ is due to spin-curvature interaction. Computing these terms for the
metric (\ref{PN-metric}), we obtained
\begin{eqnarray}
\boldsymbol{\tau}_v &&=\frac{GM}{c^2r^2} \left[ 2(\mathbf{S}\cdot  \hat{\mathbf{r}}) \mathbf{v} + (\hat{\mathbf{r}}\cdot\mathbf{v}
) \mathbf{S} - (\mathbf{S} \cdot \mathbf{v}) \hat{\mathbf{r} }\right] \, , \label{tau-v}\\
\boldsymbol{\tau}_J &&= \frac {G}{c^2} \left[\frac{3(\mathbf{J}\cdot \mathbf{\hat r}) \mathbf{\hat r}
- \mathbf{J}}{r^3} \right]\times \mathbf{S} \,, \label{tau-J} \\
\boldsymbol{\tau}_R&& = \frac{G}{c^2}\left(\frac{M}{m}\right) \left[ \frac{3(\mathbf{S} \cdot \mathbf{\hat r})
\mathbf{\hat r}}{r^3} \right] \times \mathbf{S}  \label{tau2} \,.
\end{eqnarray}
Magnitude of the torque (\ref{PN-S}) does not represent directly measurable quantity. Indeed, evolution of the
gyroscope axis is observed in the frame co-moving with the gyroscope, so the measurable quantity is $\frac{d{\bf
S}'}{ds}$, where $S'_i$ are components of spin-vector in the rest frame of gyroscope, and $s$ is its proper time.
Magnitudes of the two torques do not coincide, since ${\bf S}$ is not a covariant object. According to the classical
work of Schiff \cite{Schiff1960.1}, we can present  ${\bf S}'$ through ${\bf S}$, and then use the resulting relation to
compute $\frac{d{\bf S}'}{ds}$ through $\frac{d{\bf S}}{dt}$ given in (\ref{PN-S}). The procedure is as follows. First,
we use the tetrad formalism, presenting original metric along an infinitesimal arc of the gyroscope trajectory as
$g_{\mu\nu}=\tilde e_\mu^A\tilde e_\nu^B\eta_{AB}$. Let $e^\mu_A$ is inverse matrix of $\tilde e^A_\mu$. Applying a
general-coordinate transformation $x^\mu\rightarrow x^A$ with the transition functions $\frac{\partial x^\mu}{\partial
x^A}=e^\mu_A$, the metric acquires the Lorentz form, $\eta_{AB}=e^\mu_A e^\nu_B g_{\mu\nu}$. So the transformed
spin-tensor, $S^{CD}=\tilde e_\mu^C\tilde e_\nu^D S^{\mu\nu}$, represents spin of gyroscope  in a free-falling frame.
Second, we apply the Lorentz boost $\Lambda^C{}_A({\bf v})$, where ${\bf v}$ is velocity of gyroscope, to make the
frame co-moving with gyroscope. This gives the spin-tensor $S'^{CD}= \Lambda^C{}_A \Lambda^D{}_B \tilde e_\mu^A\tilde
e_\nu^B S^{\mu\nu}$. Then three-dimensional spin (\ref{d3s}) in the co-moving frame can be presented through the
quantities given in original coordinates as follows:
\begin{eqnarray}\label{gy1}
S'_i=\frac14\epsilon_{ijk}\Lambda^j{}_A \Lambda^k{}_B \tilde e_\mu^A\tilde e_\nu^B S^{\mu\nu}.
\end{eqnarray}
Since our metric is diagonal in $\frac{1}{c^2}$\,-approximation, the tetrad field is diagonal as well, and reads
$\tilde e_0^0=1-\frac{GM}{c^2r}$, $\tilde e_i^i=1+\frac{GM}{c^2r}$, $i=1, 2, 3$, again to $\frac{1}{c^2}$ order. The
Lorentz boost is given by the matrix with components $\Lambda^0{}_0=\gamma$, $\Lambda^i{}_0=\Lambda^0{}_i
=-\gamma\frac{v^i}{c}$, $\Lambda^i{}_j=\delta^i{}_j+\frac{\gamma-1}{{\bf v}^2}v^iv_j$, where
$\gamma=(1-{\bf v}^2/c^2)^{-\frac12}$. Using these expressions in  Eq. (\ref{gy1}), we write it in
$\frac{1}{c^2}$\,-approximation
\begin{eqnarray}\label{gy2}
{\bf S}'={\bf S}+\frac{2GM}{c^2r}{\bf S}-\frac{1}{2c^2}\left[{\bf v}^2{\bf S}-({\bf v}\cdot{\bf S}){\bf v}\right].
\end{eqnarray}
To compute derivative $\frac{d}{ds}$ of this expression, we note that the difference between $ds$ and $dt$ can be
neglected, being of order $\frac{1}{c^2}$, so we can replace  $\frac{d}{ds}$ on $\frac{d}{dt}$ on the right hand side
of (\ref{gy2}). For $\frac{d{\bf v}}{dt}$ we use  its expression (\ref{PN-F1}) in the leading approximation,
$\frac{d{\bf v}}{dt}=-\frac{MG}{r^2}\hat{\bf r}$.  The result is
\begin{eqnarray}\label{gy3}
\frac{d{\bf S}'}{ds}=&& \quad \frac{d{\bf S}}{dt} \nonumber \\
&&-\frac{GM}{c^2r^2}\left[(\hat{\bf r}\cdot{\bf v}){\bf S}+\frac12 (\hat{\bf
r}\cdot{\bf S}){\bf v}+\frac12 ({\bf v}\cdot{\bf S})\hat{\bf r}\right].
\end{eqnarray}
We substitute (\ref{PN-S}) into (\ref{gy3}), and then replace ${\bf S}$ on ${\bf S}'$ in the resulting expression,
since according to (\ref{gy2}), ${\bf S}$ differs from ${\bf S}'$ only by terms of order $\frac{1}{c^2}$. The final
result for total torque in the rest frame of gyroscope is
\begin{equation}\label{gy4}
\frac{d\mathbf{S}'}{ds} =  \boldsymbol{\tau}'_v +  \boldsymbol{\tau}'_J + \boldsymbol{\tau}'_R \, ,
\end{equation}
where
\begin{eqnarray}\label{gy5}
\boldsymbol{\tau}'_v =\frac{3GM}{2c^2r^2}[\hat{\bf r}\times{\bf v}]\times {\bf S}',
\end{eqnarray}
while $\boldsymbol{\tau}'_J$ and $\boldsymbol{\tau}'_R$ are given by (\ref{tau-J}) and (\ref{tau2}), where ${\bf S}$
must be replaced on ${\bf S}'$. \par

\noindent {\it Comments. 1.} Curiously enough, spin torque in original coordinates, being averaged over a revolution
along an almost closed orbit, almost coincides with instantaneous torque in the co-moving frame. This has been observed
by direct computation of the mean value of $\frac{d{\bf S}}{ds}$, see \cite{Thorne1998, Adler2015}. The same result is
implied by Eq. (\ref{gy3}): $\langle \frac{d{\bf S}'}{ds}\rangle - \langle\frac{d{\bf S}}{dt}\rangle\sim\frac{1}{c^2}\langle\frac{d{\bf S}}{dt}\rangle\sim\frac{1}{c^4}$, and since
$\langle\frac{d{\bf S}'}{ds}\rangle\approx\frac{d{\bf S}'}{ds}$, we have $\langle\frac{d{\bf S}}{dt}\rangle\approx\frac{d{\bf S}'}{ds}$. \par

\noindent {\it 2.} Spin-tensor subject to the condition $S^{\mu\nu}P_\nu=0$ can be used to construct four-dimensional
Pauli-Lubanski vector
\begin{eqnarray}\label{gy6}
s_\mu=\frac{\sqrt{-\det g_{\mu\nu}}}{4\sqrt{-P^2}}\epsilon_{\mu\alpha\beta\gamma}P^\alpha S^{\beta\gamma}, ~
\mbox{where} ~ \epsilon_{0123}=-1.
\end{eqnarray}
In a free theory, where $P^\alpha$ does not depend on $S^{\beta\gamma}$, this equation can be inverted, so $S^{\beta\gamma}$ and
$s_\mu$ are mathematically equivalent. Hence spatial components ${\bf s}$ could be equally used to describe spin of a gyroscope
\cite{Weinberg}. In $\frac{1}{c^2}$\,-approximation we have $P^\alpha=m\dot x^\alpha$, and (\ref{gy6}) implies ${\bf
s}={\bf S}$ in the rest frame of gyroscope. Under general-coordinate transformations, ${\bf S}$ transforms as spatial
part of a tensor, while ${\bf s}$ transforms as a part of four-vector.  So the two spins differ in all frames except
the rest frame. Let us find the relation between them in $\frac{1}{c^2}$\,-approximation. Using the approximate
equalities $(-vgv)^{-\frac12}=\frac{1}{c}\left(1+\frac{v^2}{2c^2}+\frac{GM}{c^2r}\right)$ and $\sqrt{-\det
g_{\mu\nu}}=1+\frac{2GM}{c^2r}$ together with Eqs. (\ref{gmm17}), (\ref{Ap4}) and (\ref{Ap4.1}), we obtain for spatial
part of (\ref{gy6})
\begin{eqnarray}\label{gy7}
{\bf s}=\frac{1}{\gamma}{\bf S}+\frac{1}{c^2}({\bf v}\cdot{\bf S}){\bf v}+\frac{3GM}{c^2r}{\bf S}.
\end{eqnarray}
Computing derivative of this equality and  using (\ref{PN-S})-(\ref{tau2}), we arrive at the following expression
for variation rate of ${\bf s}$:
\begin{eqnarray}\label{gy8}
\frac{d{\bf s}}{dt}=&& \frac{GM}{c^2r^2}\left[({\bf s}\cdot\hat{\bf r}){\bf v}-({\bf v}\cdot\hat{\bf r}){\bf s}-2({\bf
s}\cdot{\bf v})\hat{\bf r}\right]\nonumber \\
&&+{\boldsymbol{\tau}}_J+{\boldsymbol{\tau}}_R.
\end{eqnarray}
The first term coincides with that of Weinberg \cite{Weinberg}.

\noindent{\bf Post-Newtonian Hamiltonian.} Let us obtain an effective Hamiltonian, which yields the equations
(\ref{PN-F1}) and (\ref{PN-S}) in $\frac{1}{c^2}$\,-approximation. According to the procedure described in
\cite{DPM2016}, complete Hamiltonian for dynamical variables as functions of the coordinate time $t$ is $H=-cp_0$,
where $p_0$ is a  solution to  the mass-shell constraint (\ref{Hamiltonian-k}) with $P_\mu$ given in
(\ref{canonical-P}). Solving the constraint, we obtain
\begin{eqnarray}\label{efh1}
H=&&\frac{c}{\sqrt{-g^{00}}}\sqrt{(mc)^2+\gamma^{ij}P_i P_j+\frac{1}{16}(\theta
S)} \nonumber \\
&&-c\pi_\mu\Gamma^\mu{}_{0\nu}\omega^\nu+\frac{cg^{0i}}{g^{00}}P_i,
\end{eqnarray}
where $\gamma^{ij}=g^{ij}-\frac{g^{0i}g^{0j}}{g^{00}}$. After tedious computations, this gives the following expression
up to $\frac{1}{c^2}$\,-order:
\begin{eqnarray}\label{efh2}
H=&&mc^2+\frac{1}{2m}\left[{\bf p}+\frac{m}{c}\left(\frac{2G}{c}[{\bf J}\times\frac{{\bf r}}{r^3}]+2\frac{M}{m}\frac{G}{c}[{\bf S}\times\frac{{\bf r}}{r^3}]\right)\right]^2 \nonumber \\
&&-\frac{({\bf p}^2)^2}{8m^3c^2}-\frac{3GM}{2mc^2r}{\bf p}^2- m\frac{GM}{r}+m\frac{(MG)^2}{2c^2r^2} \nonumber \\
&&+\frac{1}{2c}\left(\frac{2G}{c}[{\boldsymbol{\nabla}}\times[{\bf J}\times\frac{{\bf r}}{r^3}]]+\frac{M}{m}\frac{G}{c}[{\boldsymbol{\nabla}}\times[{\bf S}\times\frac{{\bf r}}{r^3}]]\right)\cdot {\bf S}. \qquad
\end{eqnarray}
Together with the Dirac brackets (\ref{DB.1}), also taken in $\frac{1}{c^2}$\,-approximation, this gives Hamiltonian
equations of motion.  Excluding from them the momentum ${\bf p}$, we arrive at the Lagrangian equations (\ref{PN-F1})
and (\ref{PN-S}).

To write the Hamiltonian in a more convenient form, we introduce%
\footnote{We recall \cite{Jackson} that vector potential, produced by a localized current distribution ${\bf J}({\bf
x}')$ in electrodynamics is determined, in the leading order, by the vector of magnetic moment
${\boldsymbol\mu}=\frac{1}{2c}\int [{\bf x}'\times{\boldsymbol J}({\bf x}')]d^3 x$ as follows: ${\bf A}=[{\boldsymbol
\mu}\times\frac{{\bf r}}{r^3}]$, and the corresponding magnetic field is ${\bf B}=[{\boldsymbol{\nabla}}\times {\bf
A}]=\frac{3({\boldsymbol\mu}\cdot \hat{\bf r}) \hat{\bf r}-{\boldsymbol\mu}}{r^3}$.}
vector potential $A_{Ji}=-c^2 g_{0i}$ for the gravitomagnetic field ${\bf B}_J$, produced by rotation of central body
(we use the conventional  factor $\frac{2G}{c}$, different from that of Wald \cite{Wald1972}. In the result, our ${\bf
B}_J=4{\bf B}_{Wald}$)
\begin{eqnarray}\label{efh3}
&{\bf A}_J=\frac{2G}{c}[{\bf J}\times\frac{{\bf r}}{r^3}], \nonumber \\
 &\mbox{then} \quad {\bf
B}_J=[{\boldsymbol{\nabla}}\times {\bf A}_J]=\frac {2G}{c} \frac{3({\bf J}\cdot \hat{\bf r}) \hat{\bf r}-{\bf J}}{r^3}.
\end{eqnarray}
Then Eq. (\ref{efh2}) prompts to introduce also the vector potential ${\bf A}_S$ of fictitious gravitomagnetic field
${\bf B}_S$ due to rotation of a gyroscope
\begin{eqnarray}\label{efh4}
&{\bf A}_S=\frac{M}{m}\frac{G}{c}[{\bf S}\times\frac{{\bf r}}{r^3}], \nonumber \\
& \mbox{then} \quad {\bf
B}_S=[{\boldsymbol{\nabla}}\times {\bf A}_S]=\frac{M}{m}\frac {G}{c} \frac{3({\bf S}\cdot \hat{\bf r}) \hat{\bf r}-{\bf
S}}{r^3},
\end{eqnarray}
as well as the extended momentum
\begin{eqnarray}\label{efh4.1}
{\boldsymbol{\Pi}}\equiv{\bf p}+\frac{m}{c}({\bf A}_J+2{\bf A}_S).
\end{eqnarray}
With these notation, the Hamiltonian (\ref{efh2}) becomes similar to that of spinning particle in a magnetic field
\begin{eqnarray}
&&H=mc^2+\frac{1}{2m}{\boldsymbol{\Pi}}^2-\frac{({\boldsymbol{\Pi}}^2)^2}{8m^3c^2}-\frac{3GM}{2mc^2r}{\boldsymbol{\Pi}}^2 \nonumber \\
&& \qquad - \frac{mGM}{r}+\frac{m(MG)^2}{2c^2r^2}+ \frac{1}{2c}({\bf B}_J+{\bf B}_S)\cdot {\bf S} \label{efh5.1}\qquad \\
&&=\frac{c}{\sqrt{-g^{00}}}\sqrt{(mc)^2+g^{ij}\Pi_i \Pi_j}+\frac{1}{2c}({\bf B}_J+{\bf B}_S)\cdot {\bf S}.     \label{efh5}
\end{eqnarray}
Note that the Hamiltonian $\frac{c}{\sqrt{-g^{00}}}\sqrt{(mc)^2+g^{ij}p_ip_j}$ corresponds to the usual Lagrangian
$L=-mc\sqrt{-g_{\mu\nu}\dot x^\mu\dot x^\nu}$ describing a particle propagating in the Schwarzschild metric
$g_{\mu\nu}$. So, the approximate Hamiltonian (\ref{efh5}) can be thought as  describing a gyroscope orbiting in the
field of Schwarzschild space-time and interacting with the gravitomagnetic field.

Effective Hamiltonian for MPTD equations turns out to be less symmetric: it is obtained from (\ref{efh5}) excluding the
term $\frac{1}{2c}({\bf B}_S\cdot {\bf S})$, while keeping the potential ${\bf A}_S$ in (\ref{efh4.1}). Hence the only
effect of non-minimal interaction is the deformation of gravitomagnetic field of central body according to the rule
\begin{eqnarray}\label{efh7}
{\bf B}_J\rightarrow{\bf B}_J+{\bf B}_S.
\end{eqnarray}

\section{Discussion}\label{sect.5}

Starting from a variational problem, we have studied relativistic spinning particle with non-minimal spin-gravity
interaction through the gravimagnetic moment $\kappa$. Hamiltonian equations for an arbitrary $\kappa$ are presented in
(\ref{motion-x-k})-(\ref{motion-S-k}). When $\kappa=0$, our variational problem  yields MPTD equations
(\ref{motion-P-2}) and (\ref{motion-S}), accompanied by the momentum-velocity relation (\ref{motion-x.1}) and by the
expected constraints (\ref{primary}), (\ref{condition1}) and (\ref{condition2}). When $\kappa=1$, the MPTD equations
are modified by extra terms, see Eqs. (\ref{xk.1}) and (\ref{x.111})  above.

We have computed, in the coordinate-time parametrization $t=\frac{x^0}{c}$, the acceleration
(\ref{acc-4})-(\ref{acc-5}) and the spin torque (\ref{tau-v})-(\ref{tau2}) of our gravimagnetic particle in the field
of a rotating central body (\ref{PN-metric}) in the leading post-Newtonian approximation. We also obtained the
approximate Hamiltonian (\ref{efh5}), which implies these expressions in the Hamiltonian formulation with use of Dirac
brackets. As it should be expected, the expressions (\ref{acc-4.0}), (\ref{acc-R.0}) and  (\ref{tau-v}), (\ref{tau-J})
coincide with those of known from analysis of MPTD equations \cite{ Adler2015, Thorne1998, Einstein1915, Einstein1916,
deSitter1916, Thirring1918.2, Lense1918, Mashhoon1982, Connell1975, Connell1970, Robertson1938}. The new terms due to
the non-minimal interaction are (\ref{acc-5.0}) and (\ref{tau2}). Using the notation (\ref{efh3}) and (\ref{efh4}),
the total acceleration of spinning particle in $\frac{1}{c^2}$\,-approximation reads
\begin{eqnarray}
\mathbf{a}=&&  -\frac{MG}{r^2} \mathbf{\hat r} + \frac{4GM}{c^2r^2}(  \hat{\mathbf{r}}\cdot\mathbf{v} )
\mathbf{v} -\frac{GM}{c^2r^2} v^2\mathbf{\hat r} \nonumber \\
&&+ \frac{4G^2M^2}{c^2r^3}\mathbf{\hat r}  \label{acc-4} \\
&&+\frac{1}{c}\left({\bf B}_J+{\bf B}_S\right)\times{\bf v}+
\frac{GM}{mc^2r^3} \left[ {\bf S}\times{\bf v} \right. \nonumber \\
&& \left. +3({\bf S}\cdot(\hat{\bf r}\times{\bf v}))\hat{\bf r}\right] \label{acc-R} \\
&&- \frac{1}{2mc}{\boldsymbol{\nabla}}([{\bf B}_J+{\bf B}_S]\cdot {\bf S}).  \qquad \qquad \label{acc-5}
\end{eqnarray}
The first term in (\ref{acc-4}) represents the standard limit of Newtonian gravity and implies an elliptical orbit. The
next three terms represent an acceleration in the orbital plane and are responsible for the precession of perihelia
\cite{Einstein1915, Einstein1916, Weinberg}. The  term $\frac{1}{c}{\bf B}_J\times{\bf v}$ represents the acceleration
due to Lense-Thirring rotation of central body, while the remaining terms in (\ref{acc-R}) and (\ref{acc-5}) describe
the influence of the gyroscopes spin on its trajectory. The first term in (\ref{acc-R}) has been computed by Lense and
Thirring \cite{Thirring1918.2, Lense1918, Mashhoon1982}, the remaining terms in (\ref{acc-R}) have been discussed in
\cite{Wald1972, DPW2, khriplovich:1996}. The gravitational dipole-dipole force $\frac{1}{2mc}{\boldsymbol{\nabla}}({\bf B}_J\cdot {\bf
S})$ has been computed by Wald \cite{Wald1972}. The new contribution due to non-minimal interaction,
$\frac{1}{2mc}{\boldsymbol{\nabla}}({\bf B}_S\cdot {\bf S})$, is similar to the Wald term. The acceleration
(\ref{acc-R}) comes from second term of effective Hamiltonian (\ref{efh5.1}), while (\ref{acc-5}) comes from the last
term.

The geodetic precession  (\ref{tau-v}) comes from second term of effective Hamiltonian (\ref{efh5.1}), while the
frame-dragging precession (\ref{tau-J}) is produced by the term $\frac{1}{2c}({\bf B}_J\cdot{\bf S})$. So they are the
same for both gravimagnetic and MPTD particle. They have been first computed by Schiff \cite{Schiff1960.1}, and
measured during Stanford Gravity Probe B experiment \cite{GravityPB2011, GravityPB2015}. The term (\ref{tau2}) is due
to non-minimal interaction, and appears only for gravimagnetic particle.

Comparing the expressions (\ref{tau-J}) and (\ref{tau2}), we conclude that precession of spin ${\bf S}$ due to
non-minimal interaction is equivalent to that of caused by rotation of central body with the momentum ${\bf
J}=\frac{M}{m}{\bf S}$.

Effective Hamiltonian for the case of non-rotating central body (Schwarzschild metric) is obtained from (\ref{efh5}) by
setting ${\bf A}_J={\bf B}_J=0$. We conclude that, due to the term $\frac{1}{2c}{\bf B}_S\cdot{\bf S}$, the spin of
gravimagnetic particle will experience frame-dragging effect (\ref{tau2}) even in the field of a non-rotating central
body.

In a co-moving frame, gravimagnetic particle experiences the precession $\frac{d{\bf
S}}{dt}=[{\boldsymbol\Omega}\times{\bf S}]$ with angular velocity
\begin{eqnarray}\label{dis1}
{\boldsymbol\Omega}=\frac{3GM}{2c^2r^2}[\hat{\bf r}\times{\bf v}]+\frac{1}{2c}{\bf B}_J+\frac{1}{c}{\bf B}_S,
\end{eqnarray}
which depends on gyroscopes spin ${\bf S}$. Hence, two gyroscopes with different magnitudes and  directions of spin
will precess around different rotation axes. Then the angle between their own rotation axes will change with time in
Schwarzschild or Kerr space-time. Since the variation of the angle can be measured with high precision, this effect
could be used to find out whether a rotating body has unit or null gravimagnetic moment.

To estimate the relative magnitude of spin torques due to ${\bf B}_J$ and ${\bf B}_S$, we represent them in terms of
angular velocities. Assuming that both bodies are spinning spheres of uniform density, we write
$\mathbf{J}=I_1{\boldsymbol{\omega}}_1$ and $\mathbf{S}= I_2{\boldsymbol{\omega}}_2$, where ${\boldsymbol{\omega}_i}$
is angular velocity and $I_i=(2/5)m_ir^2_i$ is  moment of inertia. Then the last two terms in (\ref{dis1}) read
\begin{equation}\label{O-total2}
\mathbf{\Omega}_{fd}= \frac{2Gm_1r^2_1}{5c^2r^3}\left[ 3 \left([{\boldsymbol{\omega}}_1 + \rho^2
{\boldsymbol{\omega}}_2]\cdot\mathbf{\hat r}\right)\mathbf{\hat r} - ({\boldsymbol{\omega}}_1
+\rho^2{\boldsymbol{\omega}}_2)\right] \, ,
\end{equation}
where $\rho\equiv(r_2/r_1)$. Note that $\mathbf{\Omega}_{fd}$  does not  depend on mass of the test particle. The ratio
$\rho^2\equiv(r_2/r_1)^2$ is extremely small for the case of Gravity Probe B experiment, so the MPTD and gravimagnetic
bodies are indistinguishable in this experiment.  For a system like Sun-Mercury $\rho^2\sim 10^{-5}$. For a system like
Sun-Jupiter $\rho^2\sim 10^{-2}$. The new effect could be relevant to the analysis of binary pulsars with massive
companions, where the geodetic spin precession has been observed \cite{Weisberg:2002qg, Manchester:2010dh, Hotan:2004ub}.
 Besides, the two torques could have a comparable
magnitudes in a binary system with stars of the same size (so $\rho=1$), but one of them much heavier than the other
(neutron star or white dwarf). Then our approximation of a central field is reasonable and, according to Eq.
(\ref{O-total2}), the frame-dragging effect due to gravimagnetic moment  becomes comparable  with the Schiff
frame-dragging effect.

\begin{acknowledgments}
WGR thanks to Coordena\c c\~ao de aperfei\c coamento de pessoal de nivel superior (CAPES) for the financial support (Program PNPD/2017). The research of AAD was supported by the Tomsk Polytechnic University competitiveness enhancement program and by the Brazilian foundations CNPq (Conselho Nacional de Desenvolvimento Cient\'ifico e
Tecnol\'ogico - Brasil).
\end{acknowledgments}


\end{document}